\documentclass{ws-mpla}
\usepackage[super,compress]{cite}
\usepackage[breaklinks]{hyperref}  
\hypersetup{colorlinks,urlcolor=black,citecolor=black,linkcolor=black,filecolor=black}
\usepackage{breakurl}
\usepackage{graphicx}
\begin{document}

\markboth{Pablo Sanchez-Puertas \& Pere Masjuan}
{Updated pseudoscalar contributions to the hadronic light-by-light of muon $(g-2)$}

\catchline{}{}{}{}{}

\title{
UPDATED PSEUDOSCALAR CONTRIBUTIONS TO THE HADRONIC LIGHT-BY-LIGHT OF THE MUON $(g-2)$
}

\author{\footnotesize PABLO SANCHEZ-PUERTAS
}

\address{PRISMA Cluster of Excellence, Institut f\"ur Kernphysik, Johannes Gutenberg-Universit\"at\\
Mainz D-55099,
Germany\\ 
sanchezp@uni-mainz.de}

\author{PERE MASJUAN}

\address{PRISMA Cluster of Excellence, Institut f\"ur Kernphysik, Johannes Gutenberg-Universit\"at\\
Mainz D-55099, Germany\\ masjuan@kph.uni-mainz.de
}

\maketitle

\pub{Received (Day Month Year)}{Revised (Day Month Year)}

\begin{abstract}
In this work, we present our recent results on a new and alternative data-driven determination for the hadronic light-by-light pseudoscalar-pole contribution to the muon $(g-2)$. Our approach is based on Canterbury approximants, a rational approach to describe the required transition form factors, which provides a systematic and model-independent framework beyond traditional large-$N_c$ approaches. As a result, we obtain a competitive determination with errors according to future $(g-2)$ experiments including, for the first time, a well-defined systematic uncertainty.

\end{abstract}

\ccode{PACS Nos.: 12.40.-y, 13.40.Em, 13.40.Gp, 14.60.Ef.}

\section{Introduction}

The anomalous magnetic moment of the muon $a_{\mu}\equiv (g_{\mu}-2)$ has been measured up to $0.54$~ppm and is among the most precise quantities measured in particle physics, see Ref.~\refcite{Jegerlehner:2009ry} for a detailed review. The achieved precision is not only sensitive to high order quantum electrodynamics (QED) effects, but to hadronic, electroweak and --- potentially and more interesting --- new physics contributions. Given the yet negative results for new-physics  direct searches at high-energy colliders, this quantity provides an alternative and complementary tool to those searches. Indeed, there exists at present an interesting discrepancy at the $3\sigma$ level among experiment and Standard Model (SM) prediction, which reads $a_{\mu}^{\textrm{exp}} - a_{\mu}^{\textrm{SM}}=265(85)\times10^{-11}$ (c.f. Table~\ref{tab:amucont}) and would claim the existence of new physics if the hypothesis of a statistical fluctuation in the experimental result could be ruled out. 
For this reason, two new experiments at Fermilab~\cite{LeeRoberts:2011zz} and J-PARC\cite{Mibe:2010zz} have been projected aiming for a precision of around $16\times10^{-11}$, four times smaller than current experiment\cite{Bennett:2006fi}. Such precision requires though an analogous improvement on the theory side, in which the error is totally dominated from hadronic contributions, see Table~\ref{tab:amucont}. 
\begin{table}[h]
\tbl{Standard Model contributions to $a_{\mu}$.}
{\begin{tabular}{cr@{}lc} \toprule
Contribution & \multicolumn{2}{c}{$a_{\mu}\times10^{11}$} & Refs. \\ \colrule
$a_{\mu}^{\textrm{QED}}$ &  116584718&.951(80) & \refcite{Aoyama:2012wk}\\
$a_{\mu}^{\textrm{QCD}}$ & 6956&(57) & \refcite{Jegerlehner:2009ry,Davier:2010nc,Hagiwara:2011af,Kurz:2014wya,Colangelo:2014qya} \\
$a_{\mu}^{\textrm{EW}}$  & 153&.6(1) & \refcite{Gnendiger:2013pva} \\ \colrule
$a_{\mu}^{\textrm{SM}}$ & 116591826&(57) &  \\
$a_{\mu}^{\textrm{exp}}$ & 116592091&(63) & \refcite{Bennett:2006fi,Agashe:2014kda} \\ \botrule
\end{tabular}\label{tab:amucont} }
\end{table}

This situation has prompted the necessity of more precise calculations for the relevant hadronic contributions, which are the SM bottleneck in achieving an improved theoretical precision for $a_{\mu}$. This is due to two features: first, the loop integrals involved require a full hadronic description at all scales; second, such integrals sharply peak below $1$~GeV, which demands a precise description of hadronic physics in its non-perturbative regime, where perturbative quantum chromodynamics (QCD) does not apply. It is such a combination that poses a great deal for theoretical calculations.

\begin{figure}[h]
\centerline{\includegraphics[width=0.9\textwidth]{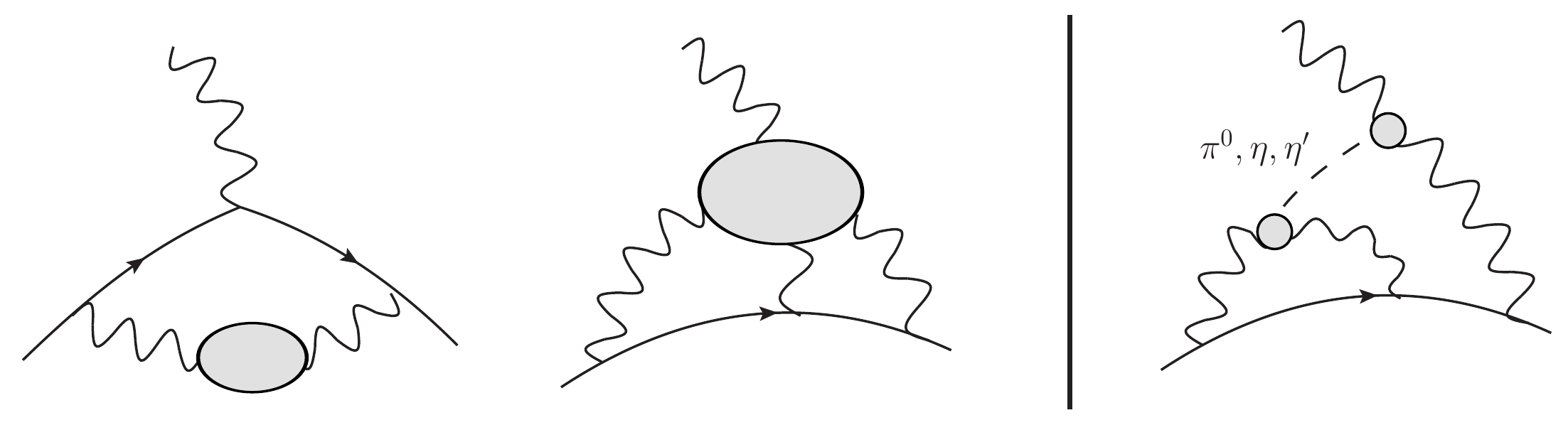}}
\vspace*{8pt}
\caption{The LO HVP (left) and the HLbL (center) contributions to $a_{\mu}$. Right: pseudoscalar-pole contribution to HLbL. \protect\label{fig:hcont}}
\end{figure}

The leading and major hadronic contribution is given by the leading order (LO) hadronic vacuum polarization (HVP), which is shown in Fig.~\ref{fig:hcont} left. Fortunately, such contribution is related through the optical theorem to the  $\sigma(e^+e^-\to\textrm{hadrons})$ cross section, which allows for a straightforward evaluation. As a consequence, increasing the current precision amounts to improve the available experimental data for the involved cross sections. 

The situation is much involved for the next-to-leading contributions since, beyond the next-to-leading-order (NLO) HVP\footnote{The NLO HVP can be easily calculated using data  along the same lines as the LO HVP\cite{Hagiwara:2011af}.}, the hadronic light-by-light (HLbL) scattering enters (Fig.~\ref{fig:hcont} center). The latter cannot be directly related to any measurable cross section and demands the knowledge of QCD at all scales, for which one needs to rely on a theoretical framework to perform such calculation. For this reason, it was devised in Ref.~\refcite{deRafael:1993za} a combined large-$N_c$ and $\chi$PT counting, allowing to split and classify the HLbL into a set of different and well-ordered contributions. Note in this respect that large-$N_c$ is the only truly perturbative approach to QCD at any scale, whereas the chiral counting allows to select those channels which are enhanced at the low energies specially relevant to $a_{\mu}$ physics. 
According to this framework, the $\pi$ and $K$ loops together with the pseudoscalars ($\pi^0,\eta$ and $\eta'$) conform the leading contributions, whereas heavier resonances and the quark-loop are subleading (further contributions can be safely neglected). The problem is reduced therefore to calculate a few contributions to the HLbL and represents the starting point for most of the HLbL calculations, which are summarized in Table~\ref{T2}.
\begin{table}[h]
\tbl{The HLbL and its contributions from different references and methods, representing the progress on the field and the variety of approaches considered. PS(HR) stands for pseudoscalar (heavier resonances) and QL for quark-loop. $\dag$ indicates used from a previous calculation.}
{\begin{tabular}{llccccl} \toprule
Authors & HLbL$\times10^{11}$ & $\pi,K$ loop & PS & HR & QL & method and year\\ \colrule
BPP~\cite{Bijnens:1995xf,Bijnens:2001cq} &$83(32)$ & $-19(13)$ &  $85(13)$ &  $-4(3)$ &  $21(3)$ & \textrm{ENJL, '95\, '96\, '02}\\  
HKS~\cite{Hayakawa:1995ps} &$90(15)$ & $-5(8)$ &  $83(6)$ &  $1.7(1.7)$ &  $10(11)$& \textrm{LHS, '95\, '96\, '02}  \\  
KN~\cite{Knecht:2001qf} &$80(40)$ & & $83(12)$ &  & &\textrm{Large $N_c$+$\chi$PT, '02}\\
MV~\cite{Melnikov:2003xd} &$136(25)$ & $0(10)$ &  $114(10)$ &  $22(5)$ &  $0$&\textrm{Large $N_c$+$\chi$PT, '04} \\    
JN~\cite{Jegerlehner:2009ry} &$116(40)$ & $-19(13)\dag$ &  $99(16)$ &  $15(7)$ &  $21(3)\dag$&\textrm{Large $N_c$+$\chi$PT, '09}\\  
PdRV~\cite{Prades:2009tw} &$105(26)$ & $-19(19)$ &  $114(13)$ &  $8(12)$ &  $0$&\textrm{Average, '09}\\  
HK~\cite{Hong:2009zw} &$107$ &  & $107$ &  && \textrm{Hologr. QCD, '09} \\  
DRZ~\cite{Dorokhov:2015psa} &$168(13)$ &  &$59(9)$ & $3(5)$ & $111(9)$& \textrm{Non-local q.m., '11} \\  
EMS~\cite{Masjuan:2012wy,Masjuan:2012qn,Escribano:2013kba} &$107(17)$ &  $-19(13)\dag$ &  $90(7)$ &   $15(7)\dag$ &  $21(3)\dag$ &\textrm{Pad\'e-data,'12\, '13}\\ 
GLCR~\cite{Roig:2014uja} &$118(20)$ & $-19(13)\dag$ &  $105(5)$ &   $15(7)\dag$ &  $21(3)\dag$& \textrm{R$\chi$T, '14} \\  \botrule
\end{tabular} \label{T2}}
\end{table}
The different results there represent different {\textit{choices}} on how to understand and {\textit{model}} the different contributions describing the relevant $\gamma^*\gamma^*M$ interactions (where $M=\pi\pi, KK, \pi^0, \eta, \eta', ... $ represents the different meson(s) involved) and briefly summarize the present status of the field. Among them, JN and PdRV represent the two standard reference numbers. More recently, different proposals appeared, among them are the lattice approaches\cite{Blum:2014oka,Green:2015sra,Blum:2015gfa}, the Dyson-Schwinger one\cite{Goecke:2010if} and the more recent dispersive approaches\cite{Pauk:2014rfa,Colangelo:2014dfa,Colangelo:2014pva}\footnote{Though the lattice and Dyson-Schwinger approaches do not fit in the scheme in Table~\ref{T2}, the dispersive approaches easily fits in such decomposition.}.

Among the different contributions in Table~\ref{T2}, it is the pseudoscalar one that dominates the HLbL and demands thereby the best precision. At present, the reference values for the latter  vary over the range of $(83-127)\times10^{-11}$, though the size of the errors yields essentially compatible results. Whereas this was an acceptable situation at the time most calculations were performed, given the present uncertainty on $a_{\mu}$, it is timely due to the future experiments precision to improve on the theoretical estimates for the HLbL. Particularly, the differences among approaches --- of the order of the projected uncertainties --- hint for non-negligible model dependencies, which along with possible systematic uncertainties must be carefully assessed. It is our purpose to update this contribution in order to meet the future experiment criteria, alleviating as much as possible previous model dependencies and unquantified systematic uncertainties. To this object, we propose to extend the framework of Pad\'e approximants (PAs) to the bivariate case. This allows to provide a model-independent description with the appropriate high-energy QCD constraints for the space-like (SL) form factors involved in the calculation.

\section{The pseudoscalar-pole contribution}
\label{sec:pspole}

\begin{figure}[h]
\centerline{\includegraphics[width=0.44\textwidth]{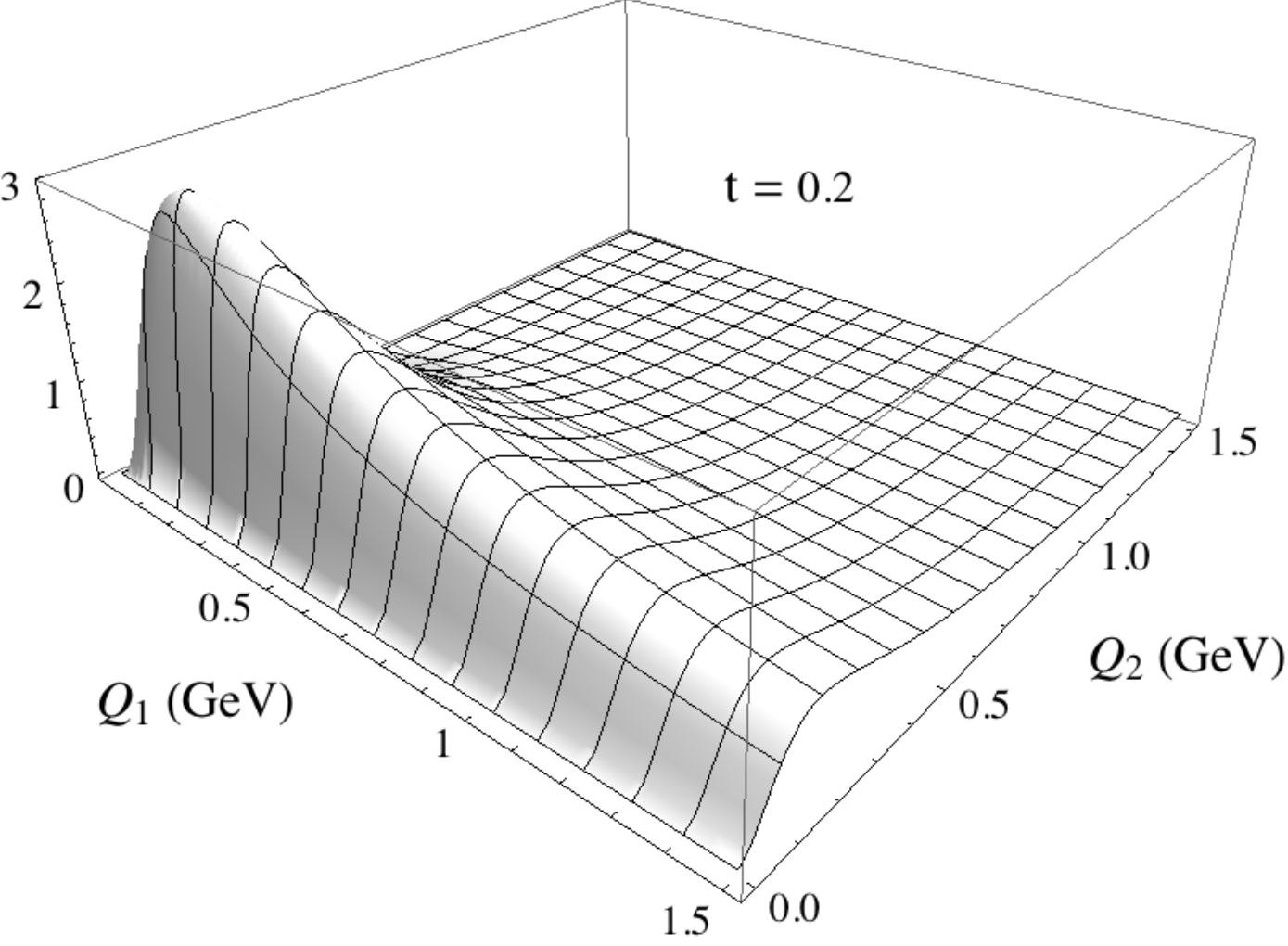},\includegraphics[width=0.44\textwidth]{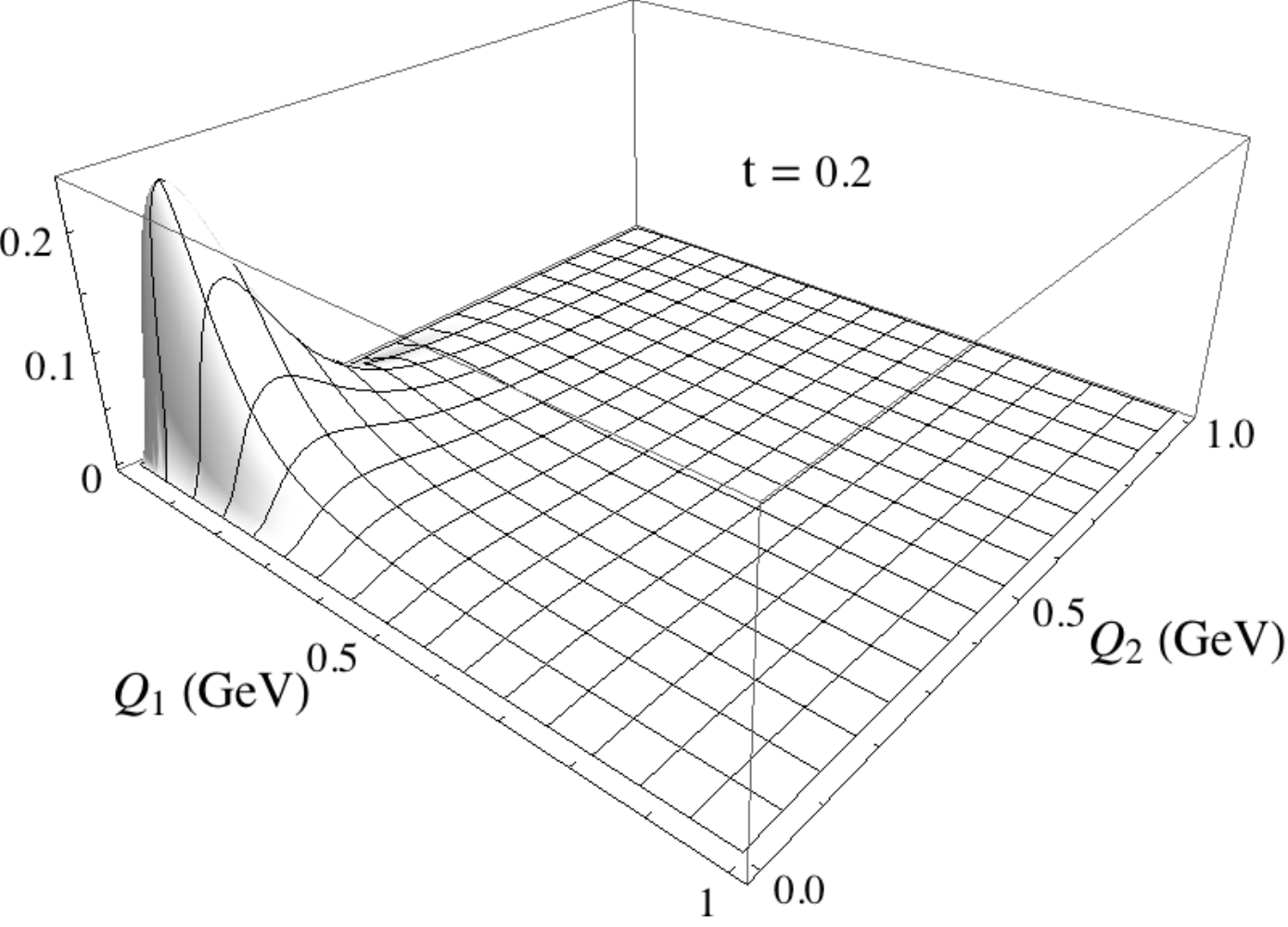}}
\vspace*{8pt}
\caption{The integrands in Eq.~\eqref{eq:HLBL} involving $I_1$ (left) and  $I_2$ (right) for $t=0.2$.\protect\label{fig:kernel}}
\end{figure}

The pseudoscalar-pole contribution to the HLbL, $a_{\mu}^{\textrm{HLbL};P}$, is depicted in Fig.~\ref{fig:hcont} right (additional permutations are implied) and involves the $P\gamma^*\gamma^*$ vertex (grey blobs in the same figure)
\begin{equation}
i\mathcal{M}_{\mu\nu} = ie^2\epsilon_{\mu\nu\rho\sigma}q_1^{\rho}\epsilon_2^{\sigma}q_2^{\sigma}F_{P\gamma^*\gamma^*}(q_1^2,q_2^2),
\end{equation}
where the pseudoscalar (on-shell) transition form factor (TFF) $F_{P\gamma^*\gamma^*}(q_1^2,q_2^2)$ appears. 
This encodes the QCD non-perturbative dynamics, and it is our ability to describe it that sets the final precision that can be reached for $a_{\mu}^{\textrm{HLbL};P}$. Explicitly, the $a_{\mu}^{\textrm{HLbL};P}$ contribution can be expressed in terms of the integral
\begin{multline}
a_{\mu}^{\textrm{HLbL};P} =  \frac{-2\pi}{3}\left( \frac{\alpha}{\pi} \right)^{3} \int_0^{\infty}dQ_1dQ_2 \int_{-1}^{+1}dt  \sqrt{1-t^2} Q_1^3Q_2^3   \\ 
                            \times \left[  \frac{F_1 I_1(Q_1,Q_2,t)}{Q_2^2+m_{P}^2}  +    \frac{F_2 I_2(Q_1,Q_2,t)}{Q_3^2+m_{P}^2}  \right]. \label{eq:HLBL}
\end{multline}
Expressions for $I_i(Q_1,Q_2,t)$ appear in Refs.~\refcite{Jegerlehner:2009ry,Nyffeler:2016gnb}~and~\refcite{Inprep} and $Q_3^2 = Q_1^2+Q_2^2 + Q_1Q_2t$, and 
\begin{equation}
F_1 = F_{P\gamma^*\gamma^*}(Q_1^2,Q_3^2)F_{P\gamma^*\gamma}(Q_2^2,0), \qquad F_2 = F_{P\gamma^*\gamma^*}(Q_1^2,Q_2^2)F_{P\gamma^*\gamma}(Q_3^2,0), 
\end{equation}
involve the TFFs, where $Q_i^2$ is a SL variable and $t$ an angular one. 
The integrands are plotted in Fig.~\ref{fig:kernel} for $t=0.2$ and a constant TFF in order to show the relevant regions in the integration. The interested reader is referred to the work in Ref.~\refcite{Nyffeler:2016gnb} and the same author's contribution to this workshop, which provides an excellent and detailed discussion about these integrals. In any case, the main features can be anticipated already from Fig.~\ref{fig:kernel}:
\begin{itemize}
   \item The $a_{\mu}^{\textrm{HLbL};P}$ contribution is sensitive to the SL region alone.
   \item Both integrands peak at low energies at around $0.1-0.2$~GeV. 
   \item The $a_{\mu}^{\textrm{HLbL};P}$ contribution is dominated by the integral involving $I_1$ (left panel in Fig.~\ref{fig:kernel}), which extends up to around $2$~GeV.
   \item The integral involving $I_1$ diverges for a constant TFF; the apppropriate high-energy TFF behavior guarantees though a finite result. 
\end{itemize}
The considerations above extend to the $\eta$ and $\eta'$ cases, with the difference that their peaks at low-energies are less pronounced, being more sensitive to the high-energy region\cite{Nyffeler:2016gnb}. 
The observations above set the requirements on the TFFs necessary for achieving a precise determination for  $a_{\mu}^{\textrm{HLbL};P}$, namely: 
\begin{itemize}
  \item A precise TFF description must be provided at very low energies as well as the region below $1$~GeV, providing the bulk of the contribution (around $90\%$, $80\%$ and $70\%$ for the $\pi^0,\eta$ and $\eta'$).
  \item The TFF must incorporate as well the appropriate high-energy behavior.
  \item An accurate prediction involves the $(1-2)$~GeV region as well.
\end{itemize}

Traditionally, the pseudoscalar TFFs have been described through large-$N_c$ based approaches\cite{Bijnens:1995xf,Hayakawa:1995ps}. However, the modelization errors in which they may incur, cannot be neglected, and are typically estimated to be of the order of $30\%$. The subsequent approach of Ref.~\refcite{Knecht:2001qf} partially circumvented these issues through the use of an hybrid large-$N_c$ data-based approach, fitting the experimental data that should reduce the model-dependency\footnote{Note that additional and more precise data for the TFFs have been released after these fits were performed\cite{Aubert:2009mc,BABAR:2011ad,Savinov:2013hda}. Accounting for these could translate into a $20\%$ shift in $a_{\mu}^{\textrm{HLbL};P}$, see Ref.~\refcite{Masjuan:2014rea}.}. However, these kind of approaches present at least two shortcomings: first, consisting in a large-$N_c$ approximation, it is dubious --- even in the SL region --- whether such approaches could reproduce the physical TFF up to an arbitrary precision; second, 
calculations carried out in the large-$N_c$ limit demand an infinite set of resonances. As such sum is not known in practice, one ends up truncating the spectral function in a resonance saturation scheme, the so-called minimal hadronic approximation\cite{Peris:1998nj}. The resonance masses used in each calculation are then taken as the physical ones from PDG instead of the corresponding masses in the large-$N_c$ limit. Both problems might lead to large systematic errors not included so far\cite{Masjuan:2007ay,Masjuan:2012wy}. Actually, a handle on the systematic error incurred could be achieved by using the half-width-rule\cite{Escribano:2013kba,Masjuan:2012sk} to ascribe $1/N_c$ corrections to the vector masses used in Refs.~\refcite{Knecht:2001qf,Nyffeler:2016gnb}~and~\refcite{Nyffeler:2009tw}. Even though this is a satisfactory way of including $1/N_c$ corrections, the final precision is not competitive with the desired accuracy goal.
It is the approach from Ref.~\refcite{Knecht:2001qf} that conforms the basics for calculating the $a_{\mu}^{\textrm{HLbL};P}$ for the two reference 
numbers,\footnote{The main difference among these numbers\cite{Prades:2009tw,Jegerlehner:2009ry} resides in their implementation of certain QCD constraints, a discussion 
which is not pursued in these contributions; for a more elaborate discussion, see Ref.~\refcite{Inprep}.} and shows the aforementioned model dependencies and the necessity 
to go beyond large-$N_c$ approaches.

It was pointed out in Ref.~\refcite{Masjuan:2007ay} that, in the large-$N_c$ framework, the Minimal Hadronic Approximation can be understood from the mathematical theory of PA to meromorphic functions. Obeying the rules from this mathematical framework, one can compute the desired quantities in a model-independent way and even be able to ascribe a systematic error to the approach\cite{Masjuan:2009wy}. One interesting detail from this theory\cite{Queralt:2010sv} is that, given a low-energy expansion of a meromorphic function, a PA sequence converges much faster than a rational function with the poles fixed in advance (such as the common hadronic models used so far for evaluating the HLbL), especially when the correct large $Q^2$ behavior is imposed.

Beyond, and more interesting, Pad\'e theory is not formally limited to the large-$N_c$ limit of QCD --- a well-known textbook example is the case of the HVP\cite{Masjuan:2009wy,Aubin:2012me} --- but can apply to the physical case (which is not possible in a resonant-like reconstruction of Pad\'e approximants). As such, it provides an excellent tool to improve upon resonant approaches and achieve a reliable value for $a_{\mu}^{\textrm{HLbL};P}$ including, for the first time, an assessment of a systematic error, which provides the model independency of the method. 

Pad\'e approximants are restricted, however, to univariate functions, whereas the $a_{\mu}^{\textrm{HLbL};P}$ involves the double virtual TFF. This requires generalizing Pad\'e theory, employed in Refs.~\refcite{Masjuan:2012wy,Masjuan:2012qn,Escribano:2013kba}, to the bivariate case, and involves the use of Canterbury approximants (CAs) described in the following section. CAs allow to implement the SL low-energy TFF behavior beyond the large-$N_c$ limitations and should be considered in this region on an equal footing as dispersive approaches.\footnote{A clear difference with respect to dispersive approaches is their ability to reproduce the resonant time-like region. Since the latter is not involved in the calculation, it is unclear whether this will introduce any gain here.} However, unlike the previous methods, CAs are not restricted to the low energies, but can be formally extended to the whole SL region which, as said, is relevant at the required precision for $a_{\mu}^{\textrm{HLbL}}$.


\section{New Approach based on Canterbury approximants}
\label{sec:cas}

Given a symmetric bivariate function, say $F_{P\gamma^*\gamma^*}(Q_1^2,Q_2^2) = F_{P\gamma^*\gamma^*}(Q_2^2,Q_1^2)$, with a known formal series expansion
\begin{equation}
\label{eq:FFexp}
F_{P\gamma^*\gamma^*}(Q_1^2,Q_2^2) = F_{P\gamma\gamma}\left[  1 - b_P\frac{Q_1^2+Q_2^2}{m_P^2} + c_P\frac{Q_1^4+Q_2^4}{m_P^4} + a_{P;1,1}\frac{Q_1^2Q_2^2 }{m_P^4}+  ... \right], 
\end{equation}
where $F_{P\gamma\gamma}=F_{P\gamma^*\gamma^*}(0,0)$, CAs\cite{Chisholm,Chisholm2,Jones:1976} are defined as rational functions of bivariate symmetric polynomials $R_N, Q_M$,
\begin{equation}
C^N_M(Q_1^2,Q_2^2) = \frac{R_N(Q_1^2,Q_2^2)}{Q_M(Q_1^2,Q_2^2)} = \frac{\sum_{i,j=0}^{N} a_{i,j}Q_1^{2i} Q_2^{2j}}{\sum_{k,l=0}^{M} b_{k,l}Q_1^{2k} Q_2^{2l}}, \quad (a_{i,j}=a_{j,i}, b_{i,j}=b_{j,i}),
\end{equation}
with coefficients $a_{i,j}$, $b_{k,l}$ defined as to match the low-energy series expansion Eq.~\eqref{eq:FFexp}, known in the mathematical jargon as the accuracy-through-order conditions\cite{BakerMorris}. For a detailed description, examples and performance of the method, the reader is referred to the Appendix of Ref.~\refcite{Masjuan:2015cjl} and Ref.~\refcite{Masjuan:2015lca}. Only in this way it is the approximant guaranteed to converge to the underlying function and accurately reproduce --- as desired --- the low-energy behavior provided it fulfills certain analytical properties\footnote{In our case, the analytical structure is unknown; the CAs practitioner has to judge {\textit{a posteriori}} if a convergence pattern is achieved or not\cite{Escribano:2015yup}.} (for instance, if the function is meromorphic\cite{Cuyt:1990} or Stieltjes\cite{Alabiso:1974vk}). Moreover, the theory formally allows to implement at the same time the high-energy behavior, allowing for a safe interpolation in the whole SL region (the resonant time-like region is out of reach in our method\footnote{See discussions in this respect in Refs.~\refcite{Escribano:2015yup}~and~\refcite{Escribano:2015nra} and possible extensions into this region in the contributions form Gonzalez-Solis in {\textit{Mod.~Phys.~Lett. A} \textbf{31}, 1630028 (2016)}.}), providing an ideal framework to describe the TFF according to the necessities outlined in the previous section.

As an example, the lowest approximant reads\cite{Masjuan:2015lca,Masjuan:2015cjl}
\begin{equation}
\label{eq:C01}
C^0_1(Q_1^2,Q_2^2) = \frac{F_{P\gamma\gamma}}{1 + \frac{b_P}{m_P^2}(Q_1^2+Q_2^2) +(\frac{2b_P^2-a_{P;1,1}}{m_P^4})Q_1^2Q_1^2}.
\end{equation}
The next approximant of interest, $C_2^1(Q_1^2,Q_2^2)$, can be schematically expressed as\cite{Masjuan:2015lca}
\begin{equation}
C^1_2(Q_1^2,Q_2^2) \!= \! \frac{a_0 + a_1(Q_1^2+Q_2^2) +a_{1,1}Q_1^2Q_2^2}{1\! +\!b_{1}(Q_1^2\!+\!Q_2^2)\! +\! b_{2}(Q_1^4\!+\!Q_2^4)\! +\! Q_1^2Q_2^2(b_{1,1}\! +\! b_{2,1}(Q_1^2\!+\!Q_2^2)\! +\! b_{2,2}Q_1^2Q_2^2)},
\end{equation}
where $a(b)_{i}\equiv a(b)_{i,0}$ and the coefficients $a_{i,j}$ and $b_{k,l}$ are related to the low-energy parameters (LEPs) in the series expansion Eq.~\eqref{eq:FFexp} (i.e., $b_P, c_P, a_{P;1,1}, ...$) via the accuracy-through-order conditions.
The knowledge of such parameters would allow to reconstruct the approximants introduced above. In general, if the full series expansion Eq.~\eqref{eq:FFexp} would be known, an arbitrary large approximant could be reconstructed. Determining as much LEPs as possible represents the main challenge and becomes the limiting factor for reconstructing the highest approximant, which finally sets the precision that can be achieved when reconstructing the TFF and, thereby, that of $a_{\mu}^{\textrm{HLbL};P}$.

Our proposal is to extract the LEPs --- not the $a_{i,j}$ and $b_{k,l}$ themselves --- in a data-driven manner, employing a fitting procedure which  has been already applied with great success in determining the $\pi^0,\eta$ and $\eta'$ single virtual TFF LEPs in Refs.~\refcite{Masjuan:2012wy,Escribano:2013kba,Escribano:2015yup}~and~\refcite{Escribano:2015nra}, obtaining their normalization, slope, curvature and third derivative. Unfortunately, there is no data yet for the double-virtual TFF, which would allow to carry out a similar exercise for extracting the required double-virtual parameters in Eq.~\eqref{eq:FFexp}, such as $a_{P;1,1}$ --- the strategy used to deal with them is outlined in the section below. It can never be overemphasized the relevance of employing the LEPs when reconstructing the approximants rather than fitting the approximants themselves. This provides the adequate reconstruction with an appropriate performance at low energies and the desired accelerated convergence. 

As a matter of proof, we have verified the performance of the method introduced above against two well-motivated theoretical models for the TFFs: a large-$N_c$ Regge model\cite{RuizArriola:2006jge,Arriola:2010aq}, and a logarithmic one\cite{Radyushkin:2009zg,Masjuan:2012wy}, where full analytical information is available. The convergence, as expected, was excellent\cite{Inprep} and did not require the use of high-order approximants. In addition, we checked that the difference among one element and the previous one provided an excellent estimation for the systematic error in $a_{\mu}^{\textrm{HLbL};P}$ calculations, which we include in the following.


\section{Pseudoscalar pole contributions to HLbL}
\label{sec:results}

In order to reconstruct the TFF, we choose the $C^N_{N+1}(Q_1^2,Q_2^2)$ sequence of approximants, that will provide the appropriate high-energy QCD constraints. As an example, $\lim_{Q^2\to\infty}C^N_{N+1}(Q^2,0)\sim Q^{-2}$, the well-known Brodsky-Lepage asymptotic behavior\cite{Lepage:1980fj}. Imposing the accuracy-through-order conditions for the single virtual coefficients involves the use of $F_{P\gamma\gamma}$ and $b_P$ for the $C^0_1(Q_1^2,Q_2^2)$ approximant, and that of $\{ F_{P\gamma\gamma}, b_P, c_P, d_P  \}$ for the $C^1_2(Q_1^2,Q_2^2)$ one\footnote{The only exception is the $\pi^0$, for what we trade $d_{\pi^0}$ (which is unknown) for the Brodsky-Lepage prediction $\lim_{Q^2\to\infty}Q^2F_{\pi^0\gamma^*\gamma^*}(Q^2,0) = 2F_{\pi}$.}. Further parameters are unknown at the moment and avoid to reconstruct the $C^2_3(Q_1^2,Q_2^2)$ approximant, setting the final precision that can be reached.
Respecting the double-virtual parameters, it is useful to recall the high-energy expansion\cite{Nyffeler:2009tw,Jegerlehner:2009ry} for the $\pi^0$ (a similar one applies to the $\eta,\eta'$)
\begin{equation}
\label{eq:ffhigh}
   F_{\pi^0\gamma^*\gamma^*}(Q_1^2,Q_2^2) = \frac{2}{3}F_{\pi}\left( \frac{1}{Q^2} - \frac{8}{9}\frac{\delta^2}{Q^4} + \mathcal{O}(Q^{-6})\right).
\end{equation}
Constraining the high-energy behavior requires then $a_{P;1,1}\to 2b_P^2$ in Eq.~\eqref{eq:C01}, and no information about the double-virtual LEPs is required at this point.\footnote{In previous references\cite{Sanchez-Puertas:2015yxa} where we reconstructed the $C^0_1(Q_1^2,Q_2^2)$ approximant alone, we employed a theoretically-motivated range $a_{P;1,1}=(0-2)b_P^2$. This was extremely important, since only the lowest element was employed. The use of an additional element in this case circumvents such problem and makes the discussion superfluous.}
The next element, $C^1_2(Q_1^2,Q_2^2)$, involves four double-virtual coefficients, $\{ a_{1,1}, b_{1,1}, b_{1,2}, b_{2,2} \}$; the high-energy behavior nevertheless requires $b_{2,2}\to0$ and only three of them need to be determined. Given the lack of information at low energies, we employ Eq.~\eqref{eq:ffhigh}, which involves two coefficients, say $F_{\pi}$ and $\delta^2$. The last double-virtual coefficient must be determined from the LEPs and involves $a_{P;1,1}$ thereby. Summarizing, $\{ a_{1,1}, b_{1,1}, b_{1,2}\}$ are related through the accuracy-through-order conditions to $\{ a_{P;1,1}, F_{\pi}, \delta^2 \}$. Still, we have to face the lack of double-virtual data that could determine $a_{P;1,1}$.\footnote{A potential source of information for $a_{P;1,1}$ and double virtual parameters are $P\to\bar{\ell}\ell$\cite{Masjuan:2015lca,Sanchez-Puertas:2015yxa} and $P\to\bar{\ell}\ell\bar{\ell}'\ell'$\cite{Escribano:2015vjz} decays.} To be as general and model independent as possible and to avoid any prejudice, we take the widest range which is allowed for it (that avoiding poles in the space-like region, a natural constraint from unitarity), obtaining a band of the kind $a_{P;1,1}\in(a_{P;1,1}^{\textrm{min}} - a_{P;1,1}^{\textrm{max}})$. In the following, we take this as an additional uncertainty in the TFF reconstruction when calculating $a_{\mu}^{\textrm{HLbL};P}$.

For the first element, the $C^0_1(Q_1^2,Q_2^2)$ approximant, the method outlined above yields $a_{\mu}^{\textrm{HLbL};P} = (64.9(3.1)+17.0(0.7)+16.0(0.6))\times10^{-11} = 97.9(3.2)\times10^{-11}$, where the different contributions refer to the $\pi^0,\eta$ and $\eta'$, respectively, and include statistical errors arising from the LEPs alone. This provides a first reasonable estimate, but entails a potentially large systematic error. Improving the latter requires to use the next element in the sequence, the  $C^1_2(Q_1^2,Q_2^2)$. Again, following the method outlined above, such element yields $a_{\mu}^{\textrm{HLbL};P} = (63.4(1.3)+16.4(1.0)+14.5(0.7))\times10^{-11} = 94.3(1.7)\times10^{-11}$, with analogous identifications as in the previous result. The systematic error that such element entails is much smaller, and can be estimated from the difference with respect to the previous one as previously explained.\footnote{To be on the conservative side, we retain the largest deviation with respect to the $C^0_1(Q_1^2,Q_2^2)$ result which is obtained within the full $a_{P;1,1}\in(a_{P;1,1}^{\textrm{min}} - a_{P;1,1}^{\textrm{max}})$ range\cite{Inprep}.} Incorporating such systematic error, we obtain as our final result\cite{Inprep}  
\begin{equation}
\label{eq:FR}
  a_{\mu}^{\textrm{HLbL};P} = 94.3(1.9)_{\textrm{stat}}(4.5)_{\textrm{sys}}[4.9]_t\times10^{-11},
\end{equation}
where the first error is statistical, arising from the LEPs and high-energy coefficients determination, the second is systematical and inherent to the $C^N_{N+1}(Q_1^2,Q_2^2)$ sequence truncation and the third is the combination in quadrature of the previous ones. Eq.~\eqref{eq:FR} represents the main result from our work; it provides a data-driven model-independent determination for the $a_{\mu}^{\textrm{HLbL};P}$ and includes, for the first time, a well-defined systematic error due to truncation. 

The obtained result can be compared to existing determinations for the pion-pole contribution, such as that from KN in Table~\ref{T2} and the {\textit{pole}} result from GLCR\cite{Roig:2014uja}, $a_{\mu}^{\textrm{HLbL};P}=82.7(2.8)\times10^{-11}$. We find an improved determination with respect to KN errors, which was not their main concern at that time. Concerning the more recent approach from GLCR, we find a non-negligible difference (i.e., close to the projected experimental error). This could be ascribed to potentially unaccounted errors inherent to the order they are working in R$\chi$PT  and the problematics in describing the $\eta-\eta'$ system, for which no data was employed there. This illustrates the concerns raised at the beginning of this manuscript with respect to large-$N_c$ approaches and the relevance of an appropriate description of the $\eta-\eta'$ system. 

As can be observed from Eq.~\eqref{eq:FR}, the achieved precision is enough to meet the future experimental errors. Still, such precision could be further improved with the release of new data. In this respect, both single-virtual and double-virtual data would be of help. Regarding this, the future data for the $\pi^0$ TFF which are being analyzed both at BES III\cite{Adlarson:2014hka} and NA62\cite{Hoecker:2016lxt}, as well as the future KLOE-2 data,\cite{Babusci:2011bg} will be very helpful.
Also the $P\to\gamma\gamma$ decays and $\eta$ and $\eta'$ TFFs measurements are welcomed. Even more interesting would be the possibility of measuring the double-virtual TFF, specially for the $\pi^0$, which could be possible in the future at BES III\cite{Adlarson:2014hka}. This would allow not only to extract double-virtual parameters, but to relax some high-energy constraints in favor of low-energy ones.

Last, but not least, our method could benefit as well from alternative approaches describing the TFFs. As an example, if the lattice (or the dispersive) community could provide a low-energy description for the double-virtual TFF, this information could be easily incorporated into our approach. Alternatively, our framework could be used to provide a sort of an analytical continuation for the TFF dispersive determinations into the high-energy space-like region. Concluding, our framework does not merely offer a competitive result for $a_{\mu}^{\textrm{HLbL};P}$, but proves as well a flexible and a complementary tool to existing approaches.

Our calculation should be understood within the framework proposed by de Rafael,\cite{deRafael:1993za} but it perfectly applies to dispersive approaches too, which do not rely on large $N_c$. Below, we include the additional contributions that needs to be incorporated on top of the pseudoscalar one. In our opinion, this requires the $\pi,K$ loops\cite{Bijnens:2001cq}, the axial resonances\cite{Jegerlehner:2015stw} and the quark-loop\cite{Bijnens:2001cq}, necessary to provide the appropriate high-energy behavior --- the scalar and tensor resonances are partially included in $\pi,K$ loops. We obtain\cite{Inprep} 
\begin{align}
a_{\mu}^{\textrm{HLbL}} &= ( 94.3(4.9)_{\textrm{PS}} -19(13)_{\pi,K} +7.5(2.7)_{\textrm{HR}} +21(3)_{\textrm{QL}} )\times10^{-11} \nonumber \\
&= 103.8(14)\times10^{-11},
\end{align}
which error is further diminished with respect to the existing reference numbers. Comparing to the reference values from PdRV and JN in Table~\ref{T2}, we find a significant reduction in the error. This is due both to the improvement on the pseudoscalar contribution as well as the more recent determination for the heavier resonances contributions employed here. Moreover, JN combine their errors linearly, whereas our combination is quadratic (similar to PdRV), which results in a further reduced error. 

\section{Conclusions}
\label{sec:concl}

To summarize, we have calculated the pseudoscalar-pole contribution to $a_{\mu}^{\textrm{HLbL}}$ within the mathematical framework of CAs. This novel method  allows to go beyond large-$N_c$ approximations and, for the first time, to provide a data-driven model-independent result for such contribution. We obtain $a_{\mu}^{\textrm{HLbL};P} = 94.3(4.9)\times10^{-11}$, which error is in accordance with future experiments. In addition, we have illustrated the advantages of our approach with respect to resonant or large-$N_c$ ones and the necessity of an accurate $\eta$ and $\eta'$ TFFs, which require accurate descriptions up to large $Q^2$ energies. Our approach fits well both in the traditional framework proposed by de Rafael 20 years ago\cite{deRafael:1993za} and the more recent dispersive one.

%

\section*{Acknowledgments}

The authors would like to thank the organizers for their organization and the nice atmosphere during the workshop. 
Work supported by the Deutsche Forschungsgemeinschaft DFG through the Collaborative Research
Center ``The Low-Energy Frontier of the Standard Model'' (SFB 1044).


\bibliographystyle{ws-mpla}
\bibliography{mybib}

%
%
%
%
\end{document}